# A Framework for Capturing and Analyzing Unstructured and Semi-Structured Data for a Knowledge Management System

Gerald Onwujekwe[1] Kweku-Muata Osei-Bryson[1]
and Nnatubemugo Ngwum[2]

[1]Department of Information Systems, Virginia Commonwealth University,
Richmond, VA, USA
[2]Department of Computer and Information Sciences,
Towson University, Maryland, USA

## Abstract

*Mainstream knowledge management researchers generally agree that knowledge extracted from unstructured data and semi-structured data has become imperative for organizational strategic decision making. In this research, we develop a framework that captures and analyses unstructured data using machine learning techniques and integrates knowledge and insight gained from the data into traditional knowledge management systems. Unlike most frameworks published in the literature that focuses on a specific type of unstructured data, our frameworks cut across the varieties of unstructured data ranging from textual data from social network sites, online forums, discussion boards, reviews to audio data, image data and video data. We highlight some preprocessing and processing techniques for these data and also highlight some standard output. We evaluate the framework by developing a textual data application programming interface (API) using python and beautiful soup and we perform sentiment analysis on the students' review data collected through the API.*

## 1. Introduction

Researchers have noted the organizational value of capturing unstructured and semi-structured knowledge. Ramesh and Tiwana [1] said that the business value of codification and capturing semi-structured knowledge was found in a recent study of 120 projects across a cross-section of firms. They further highlighted that the integration of the knowledge management system they developed with a groupware system for formal and informal interactions, is essential for a comprehensive representation of process knowledge. Knowledge management (KM) is a discipline that promotes an integrated approach to identifying, capturing, evaluating, retrieving, and sharing all the enterprise's information assets including databases, documents, and procedures, among others [2]. [3] has studied knowledge management, e-learning systems, and organizational learning as the three variables that impact organizational intelligence, with no attention paid to external sources of knowledge. Organizational intelligence which seeks to measure the ability of an organization to generate knowledge relevant to what the organization does, should not be defined based on the knowledge that could be harnessed internally within the organization only. [4] observed that some companies, like Andersen Consulting and Lotus, evaluate their employees for their annual job performance review partly on how well they contribute their knowledge to the organization's knowledge repositories and how well they use and apply the knowledge that exists in these repositories. Scanty knowledge repository is a known problem in knowledge management systems (KMS) hence incorporating external knowledge sources for specific purposes within the repository is reasonable. In this paper, we propose a framework to capture unstructured and semi-structured data for learning new knowledge patterns and discuss how the result should be integrated into a knowledge management system or business intelligence dashboard. The relevance of this framework is underscored by the fact that discussion groups, social media, and online forums are significant modes of social interaction, hence organizations cannot continue to ignore the knowledge that could be extracted from these group-interactive platforms that generate mostly unstructured data. Furthermore, researchers through the years have noted the importance of the knowledge that emanate from social interactions and social platforms and the need to harness such knowledge. [5] noted that theories of organizational learning do not address the critical notion of externalization, and organizational learning has paid little attention to the importance of socialization. The socialization mode refers to the conversion of tacit knowledge to new tacit knowledge through social interactions and shared experience among organizational members [6]. In other words, knowledge can be generated through social interactions. Verma and Singh [7] highlighted that integrating relevant information from diverse sources and utilizing it all for decision-making purposes is still a huge challenge. Informal and formal channels, such as the intranet or corporate portals, should be employed to help access knowledge [8]. We respond to this challenge raised by [7] by creating a framework that allows unstructured data from diverse sources to be processed, and insights generated from the data, stored as organizational knowledge or used for organizational decision-making. In this paper, we make a case that;

- A new paradigm of organizational knowledge should not only leverage knowledge internally within the organization but also externally among reviewers, customers, users, consumers, and other stakeholders.

- Unstructured data from non-traditional sources such as social-interactive platforms should be extracted and harnessed.

- Unstructured data that has been processed and analyzed should be stored on a knowledge management system or monitored as graphical outputs on business intelligence dashboards for interesting trends and patterns.

The rest of the paper is structured as follows; section 2 will examine the frameworks available in the literature for capturing unstructured and semi-structured knowledge and data. In section 3, we will present our framework for capturing and processing unstructured and semi-structured data. Section 4 presents the evaluation of the framework and we conclude the paper in section 5.

## 2. LITERATURE OVERVIEW

Due to the inextricable link between data and knowledge according to [9], and [10], this section of the literature review will focus on published frameworks for analyzing unstructured and semi-structured data and knowledge.

Orenga-Rogla in [11] developed a knowledge management 2.0 development framework made up of content module, transfer module, enrichment module, and decision-making module. The content module depicts are knowledge worker with tacit knowledge in the form of experience, thoughts, and competence. The transfer module components show how the knowledge worker can externalize the tacit knowledge via web 2.0 tools such as social networks, blog, and forums. The enrichment module focuses on using natural language processing and data mining to harness the knowledge while the decision-making module uses the harnessed knowledge to make decisions. While we agree with the authors that harnessing knowledge to be stored in a knowledge management systems should incorporate machine learning and data science tools which the authors used in the enrichment module, we note that the framework does not make the distinction that tools required to enrich unstructured data vary according to the nature of the data. Verma and Singh [7] developed a methodology for the integration of multi-structured data that emphasizes on generating superimposed data visualizations to facilitate interactive data exploration. The framework developed by the researchers is focused on event-driven analysis based on data captured from the web and other sources. The emphasis has been towards the retrieval of all events around a time-frame to ensure that a business analyst has a complete view of all events that happened during the period. The solution is suitable for dynamic businesses that are driven by events and quick changes in the landscape such as journalism or product marketing and not necessarily suitable for processing and storing knowledge in a knowledge management system.

In [12], Dey and others proposed a solution to integrate unstructured and structured data into enterprise analytics. In their approach, structured data was treated in the form of a time series that capture enterprise performance information such as weekly progress reports, sales figures, revenue, and stock prices, while unstructured data was taken from customer reports, reviews and feedback, discussion forums, blogs, and social media. According to [12], the framework exploits text processing and mining techniques for information extraction from unstructured sources and allow for multiple heterogeneous inputs to automate the process of knowledge discovery through correlation of information components extracted from the data. We learned some valuable lessons for our framework from the work done by [12] however, our framework is different in that we treat the various types of unstructured data uniquely, recognizing that the techniques for harnessing and processing the data will vary according to the type, a distinction they did not make in the paper.

Cheung and his colleagues in [13] developed a framework for the elicitation of knowledge from unstructured information. Their solution called multi-faceted and automatic knowledge elicitation system (MAKES) integrates the processes of collecting data, classifying unstructured information, modeling knowledge flow and social network analysis, and makes all of these actions into a connected process to audit unstructured information [14]. The system allows for retrieving, automatic classification, capturing and sharing of knowledge from unstructured information from emails, office documents, forums, bulletin boards, and blogs, which would contain multiple concepts that could be abstracted at different levels. The researchers were able to demonstrate the viability of their solution through a trial implementation and verification test conducted in the electronics industry.

## 3. PROPOSED FRAMEWORK

We propose a framework for capturing unstructured and semi-structured data and processing them to create outputs that are capable of providing insight and new knowledge patterns for the organization. Our method is supported by [15] who said that the development of the artifact should be a search process that draws from existing knowledge. Other frameworks tend to focus on one type of format, especially textual data. To the best of our knowledge, our framework is the first to incorporate all four types of unstructured data (text, image, audio, and video) in one framework.

## 3.1. Data Acquisition

The application-programming interface (API) on the framework provides data from different sources such as discussion groups, online forums, customer reviews and reports, social media, telephone conversation, mobile phone marketing, video conferencing, etc. The type of data will determine the techniques. If for instance, we extract textual data from blogs, discussion groups, and customer reviews, for preprocessing, we would apply metadata extraction, de-duplication, tagging, stemming, filtering, and stop word removal including punctuation marks and numbers. We will now walk through the techniques in the framework using textual data analytics.

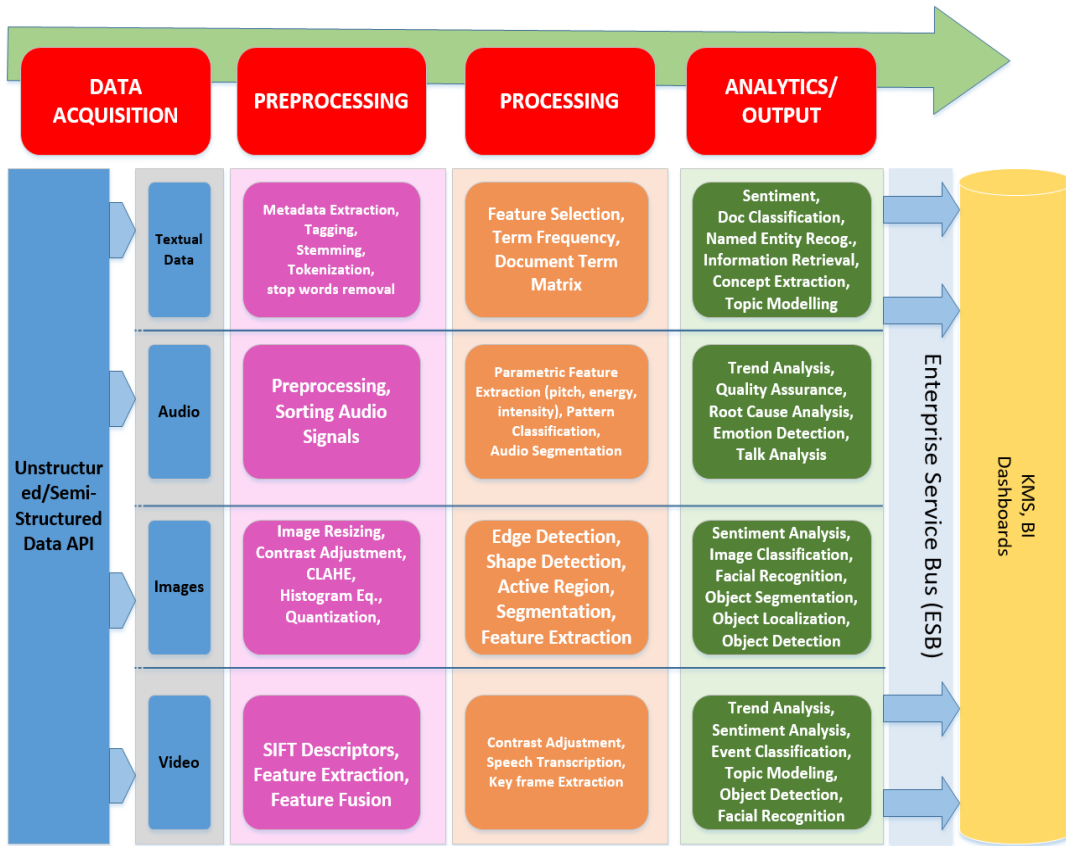

Figure 1. Proposed Framework for Capturing Unstructured and Semi-Structured Data

## 3.2. Pre-processing

*De-duplication,* also known as deduping, is a technique for removing duplicate copies and repeating textual data from the corpus. It has the benefit of reducing preprocessing and processing overhead. Textual data that contains lots of duplicates could lead to misleading output hence, it is important to apply this technique at the early stages of preprocessing.

*Metadata Extraction:* Unique properties of textual data such as the author, date, and time of creation, title, and subject may be extracted to provide relevant context during the preprocessing stage.

*Tagging:* We can apply named entity recognition (NER) or parts of speech (POS) tagging depending on the desired output. NER would allow for the identification of person names, organizations, locations, times statement, etc., that is present in the data while POS tagging will allow for classifying words and labeling them accordingly.

*Tokenization:* Given the textual data under preprocessing, we apply the tokenization technique to split up the text into individual words, phrases, or whole sentences.

*Stemming* returns every derived word to its root or base form. This densifies the data, reduces the number of words used in the corpus, and results in more efficient processing of corpus.
*Lemmatization* procedure works similarly to stemming; however, it uses a dictionary to ensure that the derived word returns to the dictionary or base form.

*Stop Words:* Prepositional words, conjunctions, and commonly used words are removed from the corpus using the stop words removal procedure. So also, irrelevant numbers such as page numbers and punctuation marks.

### 3.3. Processing

*Feature Selection* technique allows us to select a smaller group or subset of terms in the corpus and using only this subset in further processing and analysis. This increases the efficiency of the algorithm by decreasing the vocabulary lexicon.

*Term Frequency* calculation gives the number of occurrences of a given term in the textual document. There are various term frequency calculations we could use but it depends on the objectives of the analytics. Examples include simple term frequency tf(term, document), inverse document frequency (idf), combined term frequency-inverse document frequency (tf-idf), term frequency adjusted for total document length (term count/total term count in document) and augmented frequency (term frequency/highest term frequency in document).

*Document Term Matrix (DTM)* calculates the term frequency per document and creates and (m by n) matrix where m is each document in the corpus and n is the terms in the document.

### 3.4. Analytics/Output

*Sentiment Analysis:* Sentiment analysis could be applied to textual data from sources such as discussion groups, online forums, customer reviews, and reports and social media to determine the overall attitude of users, customers, etc. to a product or service that an organization creates. The knowledge derived from the analysis could provide an early warning or inform the strategy for new product development.

*Concept Extraction* technique results in the extraction of concepts from the text. A common technique for concept extraction is the use of Word Tree. When certain terms occur side by side very frequently, there is a probability that it may be alluding to a concept.

*Classification* techniques will be used to manage, sort, and group textual data in predefined categories to increase information discovery and make all discovered knowledge available and useable to support decision-making.

*Document Classification* is a technique that is used to determine the major subject or theme of a document and then assign the document to predetermined classes or categories. Document classification is a useful technique in library science and information science for document management.

*Information Retrieval* is a technique that is used on the web, knowledge management systems and other types of information systems to ensure that the right information is retrieved in the form of a search result.

*Topic modeling* is used to find hidden or abstract topics that are embedded in a collection of textual documents.

### 3.5. Integration with Traditional KMS and Business Intelligence (BI) Dashboards

We propose an Enterprise Service Bus (ESB) as the middleware for integrating analytics output with the knowledge management system.

An Enterprise Service Bus (ESB) combines event-driven and service-oriented approaches to simplify the integration of business units, bridging heterogeneous platforms, and environments [16]. An ESB is an ideal backbone for implementing service-oriented architectures because it provides a universal mechanism to interconnect all the services fully integrated business solutions without compromising security, reliability, performance, and scalability [17]. It supports synchronous and asynchronous, facilitating interactions between one or many stakeholders - one-to-one or many-to-many communications [16]. Industry-standard enterprise service bus solutions include IBM WebSphere, Oracle ESB, and SAP PI. We argue that insights and new knowledge gained from analyzing these data should be integrated into a traditional knowledge management system to enhance the richness of the knowledge available in the KMS to improve organizational performance and decision-making. Organizations often use business intelligence dashboards to monitor key performance indicators and other relevant metrics drive the business. A dashboard provides a rich user interface that displays the information in a graphical form using a variety of elements including charts, tables, and gauges. These elements reduce the time spent on analyzing the data using databases and thus assist in automating the business decision-making process [18]. While analytics results should be stored in a KMS, we argue that relevant analytics that is used in decision making should be placed on business intelligence dashboards for continuous monitoring. As such we provide our evaluation results in the form of graphics that could also be displayed in business intelligence dashboards.

### 4. FRAMEWORK EVALUATION

The effectiveness of an artifact such as a framework should be determined using an established design science evaluation method. [19] and [20] list the proven methods for evaluating a design science artifact. We prefer the more recent [20] that lists evaluation method types to include Logical Argument, Expert Evaluation, Technical Experiment, Subject-based Experiment, Action Research, Prototype, Case Study, and Illustrative Scenario. It defines Illustrative Scenario as the application of the artifact to a synthetic or real-world situation aimed at illustrating the suitability or utility of the artifact. We use the Illustrative Scenario as our method of evaluating the framework because it is the most suitable method given the nature of the artifact and the time available to complete the evaluation. We acknowledge that Case Study and Action Research evaluation methods may be more robust methods to use but they would take considerably more time to complete.

To validate the framework, we use Python and Beautiful Soup library as API to crawl textual data. An Application Programming Interface (API) has been defined as a specification that defines an interface for Software components to communicate with each other [21]. In this scenario, the specification is the HTML protocol and the software components are the website and python IDE. As highlighted earlier, each of the separate types of unstructured data in our framework may require separate procedures for evaluation. [19] highlighted that the evaluation of a designed IT artifact requires the definition of appropriate metrics and possibly the gathering and analysis of appropriate data. As such, we focus on textual data for our evaluation. We choose www.studentsreview.com and we scrap ten students' reviews of the institution where this research was performed. We name the students student_1 to student_10.

We write our API programming code as follows:

```python
def api(url):
    '''Returns student review from studentreviews.com.'''
    page = requests.get(url)
    sp = BeautifulSoup(page.text, 'html.parser')
    text = [p.text for p in sp.find(class_="portfolioContainer")]
    return text
```

### 4.1. Pre-processing

For our purposes, we will remove punctuations, brackets, parentheses, and common English stop words. We do not need to apply the stemming processing.

```python
def clean_corpus(text)
    text = text.lower()
    text = re.sub('im', '', text)
    text = re.sub('ive', '', text)
    text = re.sub('got', '', text)
    text = re.sub('isnt', '', text)
    text = re.sub('\[.*?\]', '', text)
    text = re.sub('\w*\d\w*', '', text)
    return text
```

### 4.2. Processing

The purpose of the text analytics determines the type of processing we apply to the text corpus. Based on our framework, text analytics can be performed for text classification, concept extraction, sentiment analysis, topic modeling, named entity recognition, event extraction, and information retrieval. For our evaluation, we apply document-term matrix to allow us to perform word frequency analysis for each student. We build a word cloud and show the level of profane words used by each student. The plot for each processing step is shown in the figures below:

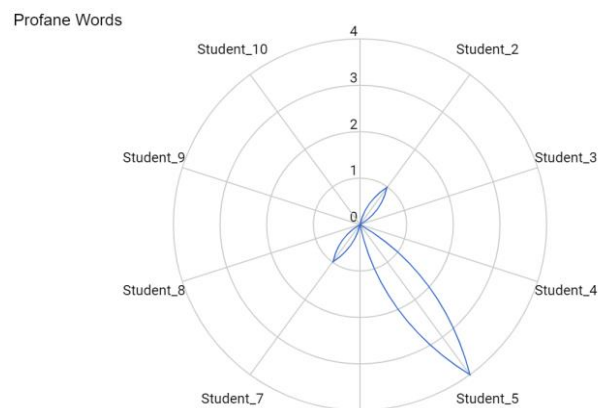

Figure 2. Profane Words Radar Chart

In the profane word radar chart, we show the level of profane words that each student is using. In our code description, profane words include words such as hell, f*ck, assh*le. The profane word analysis reveals that student_5 uses the most profane words with a total of four words. Student_2 and student_7 used one profane word each in their review while the rest of the students did not use profane words.

We show a word cloud to see the words that the students used most frequently in the reviews. From the word cloud, we see that some students focused their reviews on the school as a whole, while others on their major, school clubs, social activities, and even kids.

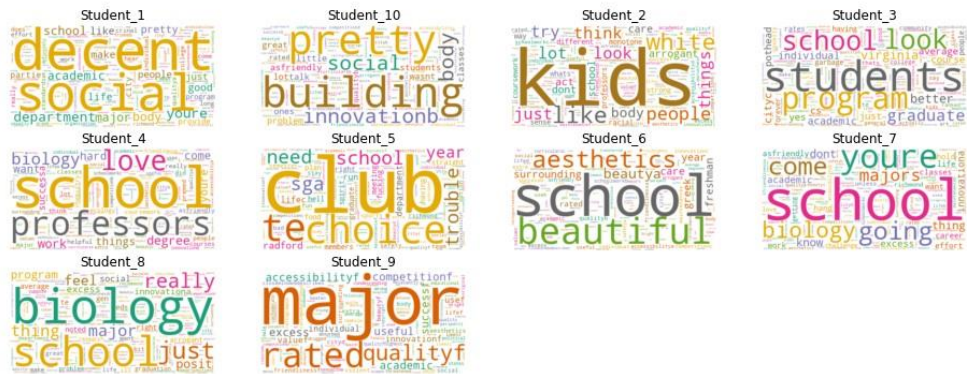

Figure 3. Word Cloud showing most frequent words for the students

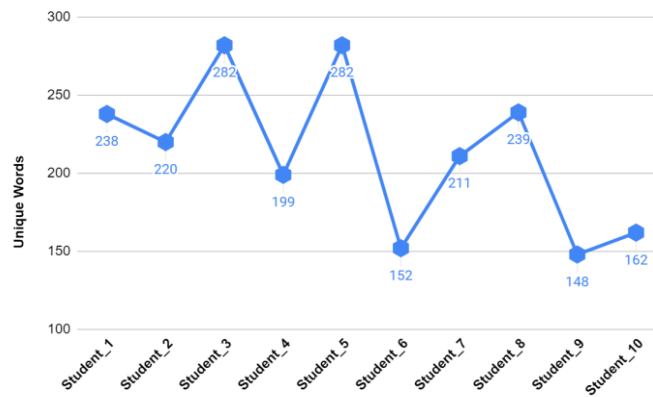

Figure 4. Unique Words of each student

The unique words plot shows the number of unique words a student used that none of the other students in the review used. The unique word plot is an indication of the length and detail of the review. It is easy to run a simple word count that would tell the length of the review however, unique words are better in this scenario because some words such as faculty, course, class, classroom, campus, school, etc. will occur across all the reviews, hence unique word count is a better indicator of how lengthy and detailed the review is. Unique word count is also an indicator of the unique points the student makes in the review. A review with a high number of unique words is expressing more views and touching on more points than one with a lesser number of unique words. The result showed that student_3 and student_5 had more to say that is different from what every other person is saying while student_9 has the least to say that is different from what others have said.

## 4.3. Analytics Type

The analytics type that fits the most based on the text corpus is sentiment analysis. Sentiment analysis usually relies on applying machine learning techniques to classify texts based on a collection of features extracted from the text using Natural Language Processing techniques, such as the presence of certain words or the coverage of some topics [22]. Topic modeling does not fit the corpus because we already know what the topic is – student giving their opinion and experience about the university they graduated from. Other types of text analytics such as concept extraction, document classification, and named entity recognition also are not the best fit given our context and nature of corpus.

The sentiment analysis plot is shown in figure 5. The horizontal axis represents the polarity of the sentiment while the vertical axis represents the subjectivity or opinion level of the review. For the polarity, zero represents a generally neural review, values less than zero represent a negative review overall, and values greater than zero represent a positive review overall. An objective sentence expresses some factual information about something, while an opinion sentence expresses some personal opinions, beliefs, feelings, allegations, desires, suspicions, and speculations. Objectivity and opinion are opposites hence a review that is high in opinion is automatically low in objectivity and vice versa. Student_7 scored the least in opinion which on the other hand means that student_7 presented the most objective review. Student_10 presented the most opinionated review which in turn means that student_10 review is the least objective. The polarity of our analysis ranges from -0.03 for student_2 to +0.21 for student_6. Student_6 submitted the most positive review about the school and student_2 submitted the most negative review.

The sentiment analysis time series for all 10 students are shown in figure 6. The vertical axis represents the polarity of the review and the horizontal axis is the timeline from 0 to 10, with zero representing the starting sentence in the review and ten the last sentence. The orange horizontal line is the neutrality line. This plot shows how the sentiment of the students varies across the entire length of the review. Some students stayed entirely positive in their review such as student_4 and student_6 which is an indication that they felt entirely positive about their experience at this university. Most other reviewers were swinging from positive to negative sentiments with student_2 having an overall negative polarity.

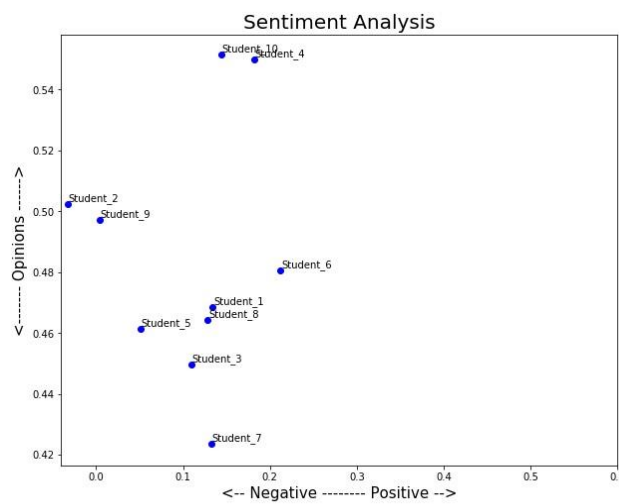

Figure 5. Sentiment Analysis

### 4.4. Integration to a Knowledge Management System

The sentiment analysis plot can be stored in an organizational knowledge management system and reviewed periodically based on the stipulations of the organizational policy. The graph analysis could also be integrated into a business intelligence dashboard and reviews with extreme negative or positive polarity could be further investigated to find what the student liked the most or hated the most about their experience. Ideas and patterns that are repeating across multiple students could inform future decisions to improve college experience for students.

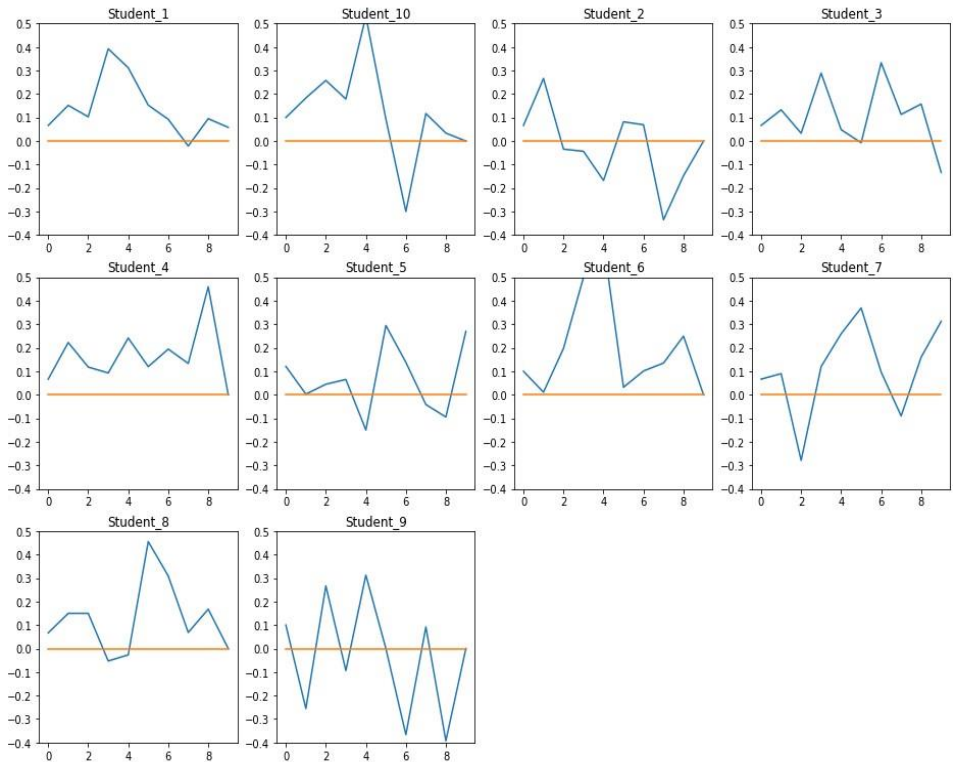

Figure 6. Sentiment Analysis Time Series for each student

## 5. CONCLUSIONS

In this paper, we have developed a framework to process and unstructured data and knowledge from multiple sources and incorporate the process output into a traditional knowledge management system. We note that the framework would be infeasible to evaluate by one straightforward procedure, but rather would require that specific instantiation be made from the framework and evaluated. In evaluating the framework, we leaned on the design science research evaluation methods provided by [20] and we used the Illustrative Scenario method. An instance of the framework was created that used student reviews from studentsreview.com to perform sentiment analysis. We used python and beautiful soup as API to crawl student review data from the website and we performed unique word count, profane word count, word cloud, sentiment analysis, and sentiment analysis time series. We propose that the result from the sentiment analysis and sentiment analysis time series could be stored in a knowledge management system and reviewed from time to time or could form an input to a business intelligence visualization tool for real-time monitoring. For future research, we hope to develop a software application interface that can sentiment-analyze user reviews from several social network feeds.

**AUTHORS**

**Gerald Onwujekwe** is a Ph.D. candidate in Information Systems at Virginia Commonwealth University, Virginia, USA. His research interests include Graph Databases, Convolutional Neural Networks, and Deep Learning and Text mining.

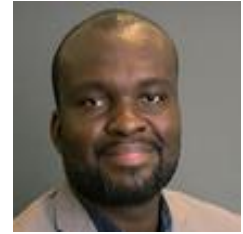

**Kweku-Muata Osei-Bryson** is a professor of Information Systems at Virginia Commonwealth University and Research Fellow of the Information Systems Research Institute since Fall 1998. Previously I was Professor of Information Systems and Decision Analysis in the School of Business at Howard University, Washington, DC, U.S.A. I have also worked as an Information Systems practitioner in both industry and government. Analytics & Data Science, Data Mining, Cyber-Security, Knowledge Management, ICT for Development, Expert & Decision Support Systems, ICT & Productivity, Multi-Criteria Decision Analysis.

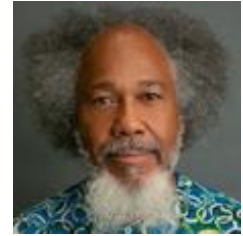

**Nnatubemugo Ngwum** is a doctoral candidate at Towson University, Maryland, USA. Obtaining a master's degree from the University of Manchester (UK) in computer security, his research interests extend beyond information systems and their security to include Internet of Things (IoT), Internet of Things security, and Human-Computer Interaction.

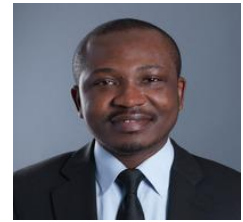